# Methanol Formation via the Radical-radical Reaction of OH and CH$_3$ Radicals Undergoing Transient Diffusion on Ice at 10 to 60 K

Arisa Iguchi[1], Hiroshi Hidaka[1*], Atsuki Ishibashi[2], Masashi Tsuge[1], Yasuhiro Oba[1], Naoki Watanabe[1]

[1]Institute of Low Temperature Science, Hokkaido University, N19W8, Kita-ku, Sapporo, Hokkaido 060-0819, Japan
[2]Komaba Institute for Science and Department of Basic Science, The University of Tokyo, Meguro, Tokyo 153-8902, Japan
*e-mail: hidaka@lowtem.hokudai.ac.jp

Abstract

Methanol (CH$_3$OH) is thought to form on interstellar ice dust via successive hydrogenation reactions. The reaction between CH$_3$ and OH radicals could also conceivably generate methanol at temperatures above approximately 20 K, at which temperature hydrogen atoms will not adhere to the ice surface. However, this process has not been verified by well-controlled experiments. Using a newly-developed Cs$^+$ ion pickup technique, the authors investigated the reaction between CH$_3$ and OH radicals on the surface of amorphous solid water, an ice dust analogue, at temperatures from 10 to 60 K. In the present experiments, OH radicals were generated by UV photolysis of water molecules, following which methane (CH$_4$) gas was deposited on the ice substrate. The results show that CH$_3$OH was formed on the ice surface through the sequential reactions CH$_4$ + OH → CH$_3$ + H$_2$O and CH$_3$ + OH → CH$_3$OH even at 10 K. Considering the very low surface coverage of reactants in the experimental condition, the second reaction was found to occur as a result of transient diffusion of CH$_3$ due to the heat of the first reaction.

# 1. INTRODUCTION

With advancements in astronomical observation technologies, many different molecules, including complex organic molecules (COMs), have been identified in association with various astronomical targets (McGuire 2018; Sakai et al. 2018; Rocha et al. 2024). It is important to assess the origin of interstellar COMs as a means of understanding the initial evolution of planetary systems. It is generally accepted that reactions on ice dust play crucial roles in producing COMs (e.g., Herbst 2017), and these reactions can be divided into energetic and non-energetic processes. Many experimental studies have demonstrated that energetic processes, specifically UV photolysis and the bombardment of ice dust by cosmic rays, can generate COMs such as amino acids (e.g., Kobayashi et al. 1995; Bernstein et al. 2002; Muñoz Caro et al. 2002; Elsila et al. 2007), sugars (Meinert et al. 2016), and nucleobases (Oba et al. 2019). In contrast, non-energetic processes not involving external energy inputs may be the primary contributors to molecule formation in dense, low-temperature regions with weak UV fields. In the case of these non-energetic processes, the chemical reactions occurring on ice dust can be categorized by temperature. At temperatures below approximately 20 K, hydrogen atom reactions dominate the surface chemistry because only hydrogen atoms can efficiently migrate over the ice surface to encounter reaction partners. As an example, it is well known that the primordial compounds water, ammonia, methane, formaldehyde, and methanol were generated in abundance by successive hydrogenation reactions occurring on ice dust (e.g., Ioppolo et al. 2008; Miyauchi et al. 2008; Tsuge et al. 2024; Hidaka et al. 2011; Qasim et al. 2020; Hiraoka et al. 1994; Watanabe & Kouchi 2002). At dust temperatures above approximately 20 K, hydrogen atoms do not remain on the ice surface for prolonged durations and so hydrogenation reactions are significantly suppressed. Instead, heavier species start to migrate and encounter one another. Under such conditions, the barrierless reactions of radicals diffusing are thought to efficiently produce various COMs (e.g., Hollis & Churchwell 2001).

Garrod *et al*. (2008) developed a chemical model for such processes that incorporated various radical reactions on ice. Their work indicated the importance of radical reactions within relatively warm regions such as hot cores and corinos in terms of contributions to chemical complexity. Experiments using infrared spectroscopy and/or temperature-programmed desorption (TPD) have been performed to confirm the formation of COMs through radical reactions (Fedoseev et al. 2015; Ioppolo et al. 2021; He et al. 2022; Santos et al. 2022). However, the conditions employed in such experimental work cannot fully reflect the phenomena occurring on ice dust, because the surface amounts of reactants were extremely high in the experimental studies compared with those on actual dust. Furthermore, the radical intermediates were not directly identified in situ during these experiments.

In recent years, COMs have been detected even in cold molecular clouds, in which the diffusion of heavy radicals over the dust surfaces should be limited (Vastel et al. 2014; Jiménez-Serra et al. 2016; Soma et al. 2018). These observations suggest that additional non-thermal diffusion mechanisms may occur at low temperatures, although COMs formation mechanisms not involving surface diffusion have also been suggested (e.g., Fedoseev et al. 2015; Chang & Herbst 2016; Simons et al. 2020; Jin & Garrod 2020). Unfortunately, radical diffusion and reaction mechanisms on ice are not as well understood as the mechanisms associated with hydrogenation reactions. To better understand the origin of COMs, it would therefore be helpful to clarify the behaviors of radicals on ice dust (e.g., Tsuge & Watanabe 2023). Studying the behaviors of radicals on ice surfaces will require highly sensitive in-situ surface-selective techniques that outperform conventional techniques such as infrared spectroscopy and TPD.

Recently, the authors developed a highly sensitive, non-destructive analytical technique referred to as $Cs^+$ pickup (Ishibashi et al. 2021). This is a powerful tool for the in-situ analysis of radicals present on ice at low amounts. Using this method, we confirmed the formation mechanism of methylformate, $HC(O)OCH_3$, on ice at 10 K.

For the methanol formation processes, successive CO hydrogenation has been considered as a dominant mechanism below approximately 20 K. Recently, interactions of $CH_3^+$ ions with ice was reported to produce methanol (Nakai et al. 2023). In addition to these processes, it has been suggested that the radical–radical association reaction ($CH_3 + OH \rightarrow CH_3OH$) is a possible process on ice even at relatively high temperatures in which CO hydrogenation is suppressed. This radical-radical association reaction has been incorporated into chemical models (Garrod et al. 2008) because parent radicals can be readily produced through the hydrogenation of C and O atoms and the photodissociation of $CH_4$ and $H_2O$. On this basis, the present study examined methanol formation via the reaction of $CH_3$ and $OH$ radicals on ice at temperatures in the range of 10–60 K.

## 2. EXPERIMENTS

Experiments were performed using the previously developed $Cs^+$ pickup technique. Details of the experimental procedure have been described elsewhere (Ishibashi et al. 2021, 2024a). Throughout the experiments, the main chamber was held at a base pressure of less than $1.0 \times 10^{-8}$ Pa. Amorphous solid water (ASW) was prepared (as an analogue for ice dust) on an aluminum substrate at 30 K via background deposition of $H_2O$ vapor at a pressure of $1.0 \times 10^{-5}$ Pa for 10 minutes. This process was estimated to deposit the ASW at a thickness of approximately 20 monolayers. The OH radicals were produced by exposing this ASW to UV radiation at approximately $6 \times 10^{12}$ photons $cm^{-2}$ $s^{-1}$ using a standard deuterium lamp at the

115–400 nm range for 1–10 minutes. The major photodissociation path under these conditions generated H + OH, although processes generating $H_2$ + O or 2H + O also likely occurred (Slanger & Black 1982). At 30 K, volatile photoproducts such as H and $H_2$ would be expected to immediately desorb from the ASW after being formed. The extent of OH coverage of the ASW surface was estimated to be on the order of 0.01 (corresponding to approximately $10^{13}$ molecules $cm^{-2}$) after 10 minutes of irradiation, based on the UV photon fluence and the photodissociation cross-section of $H_2O$ (Slanger & Black 1982). To simplify the calculations, this coverage value was assumed during the subsequent analysis.

After UV irradiation, the ASW was cooled to 10 K and gaseous $CH_4$ (Takachiho Chemical Industrial Co. Ltd., purity 99.999%) was deposited onto the sample surface at a flux of $1.4 \times 10^{11}$ molecules $cm^{-2}$ $s^{-1}$ for 15 minutes using a dedicated gas line ending in a microcapillary plate. The resulting coverage of $CH_4$ was estimated to be approximately 0.13. During $CH_4$ deposition, $CH_3$ radicals were produced as the reaction of $CH_4$ + OH → $CH_3$ + $H_2O$ took place on the ASW surface.

The $Cs^+$ pickup technique allowed non-destructive monitoring of the reactants and products during the present work, without itself promoting any undesired reactions (Ishibashi et al. 2021). In this process, $Cs^+$ ions ($m$ = 133 u) having low energies (typically approximately 17 eV) are injected onto the sample surface, upon which some these ions form complexes with neutral molecules adsorbed on the ice. The ionic complexes are scattered from the ice surfaces and subsequently assessed using a quadrupole mass spectrometer without an ionization cell.

At low coverages, the raw intensity of each pickup signal (in units of counts per second) is proportional to the surface number density of the adsorbed compound. Consequently, the adsorbate pickup efficiency ($\varepsilon_X$), defined as the ratio of the raw signal intensity ($I_{Raw[X]_t}$) to the surface number density ($[X]_t$), is an important parameter. It should be noted that $\varepsilon_X$ is significantly influenced by the $Cs^+$ flux, the temperature of the sample, and its surface condition. Even in the case of experiments conducted under the same conditions, slight differences in the $Cs^+$ flux and the microscopic surface structures will affect the pickup efficiencies. Therefore, the raw signal intensity was normalized relative to that of $H_2O$ ($I_{Raw[H_2O]_t}$), which is typically $10^4$ counts per sec. Hereafter, the normalized intensity is referred to as the "pickup intensity" ($I_{[X]_t} = I_{Raw[X]_t}/I_{Raw[H_2O]_t}$). By introducing this concept, at a given temperature, the value of relative pickup efficiency $\varepsilon'_X$ for a given adsorbate relative to that for $H_2O$, $\varepsilon_X/\varepsilon_{H_2O}$, can be assumed to be constant regardless of the $Cs^+$ beam and surface conditions. For this reason, the raw signal intensities were normalized relative to the value for $H_2O$.

## 3. RESULTS AND DISCUSSION

*3.1 Formation of CH$_3$OH following CH$_4$ deposition on OH-adsorbed ASW at 10 K*

Figure 1(a) shows the pickup spectra obtained from a pure ASW sample prepared by the background deposition of H$_2$O (in black) and from another ASW after UV irradiation for 10 minutes at 10 K (in red). As shown in the inset, both pickup spectra primarily consist of a series of H$_2$O signals, including multiple H$_2$O pickup events, appearing at 151 u (133 u + 18 u), 169 u, and 187 u. Additionally, adsorbates that had been present in very low amounts were clearly detected. These included H$_2^{18}$O (with a natural abundance of approximately 0.2% relative to the amount of H$_2^{16}$O) at 153 u, with a reasonable intensity consistent with the expected isotopic abundance. A comparison of the pickup spectra in Figure 1(a) indicates the formation of OH radicals through UV photolysis based on the signal at 150 u. The less intense peak at 167 u most likely originated from H$_2$O$_2$ formed by the recombination of OH radicals (Yabushita et al. 2008). In addition, the small peak at 168 u was attributed to the simultaneous pickup of a single OH radical and H$_2$O molecule. Figure 1(b) presents the pickup spectra acquired after 15 minutes of CH$_4$ deposition on pure ASW (in black) and UV-processed ASW (in red). Signals originated from CH$_3$ and CH$_3$OH appear solely in the latter spectrum, at 148 u and 165 u, respectively. This spectrum also shows a significant depletion of OH radicals at 150 u as compared to the red trace in Figure 1(a). These results confirm that both CH$_3$ radicals and CH$_3$OH were generated by the reaction of CH$_4$ with OH radicals on the ASW. Although C$_2$H$_6$ is a plausible product of the CH$_3$ + CH$_3$ recombination reaction, no pickup signal was detected at 163 u. This is likely due to the very low encounter probability between CH$_3$ radicals, which results from their extremely low surface coverage. Under such conditions, detecting C$_2$H$_6$ becomes even more difficult because of its low pickup efficiency.

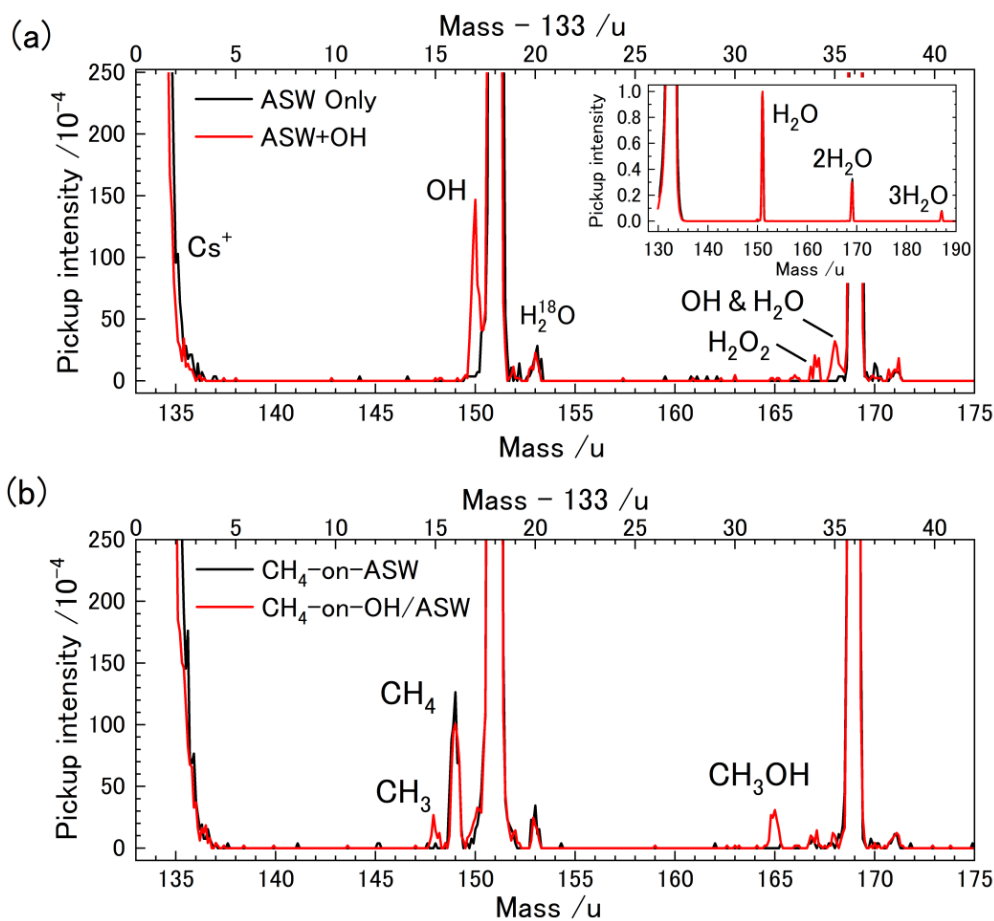

**Figure 1.** Pickup spectra obtained from (a) pure ASW (black) and ASW following UV irradiation (red) and from (b) ASW following deposition of $CH_4$ (black) and ASW following UV irradiation and deposition of $CH_4$ (red). The spectra were measured at 10 K. The UV irradiation time was 10 min, corresponding to an OH coverage of approximately 0.01, whereas the $CH_4$ deposition time was 15 minutes (coverage 0.13). The inset to (a) shows expanded mass spectra over the mass range from 128 to 190 u. The lower horizontal axis in each spectrum represents the total ion mass (including $Cs^+$ (133 u) and the adsorbate molecule) whereas the upper axis represents the mass number of only the adsorbate.

The possible chemical reactions between $CH_4$ and OH, along with their corresponding gas phase reaction energy, were calculated by Tao & Li (2002). Among these reactions, only one pathway is exothermic: the formation of $CH_3$ radical via H abstraction by OH, as shown below:

(a) $CH_4 + OH \rightarrow CH_3 + H_2O$,       $-55.2$ kJ mol$^{-1}$ ($-0.572$ eV)

This reaction has an activation barrier of approximately 26.8 kJ mol$^{-1}$ (0.278 eV), corresponding to approximately 3220 K (Espinosa-Garcia & Corchado 2015). It has been demonstrated, experimentally and theoretically, that this reaction proceeds via quantum mechanical tunneling in a low temperature ice (Lamberts et al. 2017; Zins et al. 2012). Thus, it is reasonable to consider that this pathway is the source of the CH$_3$ radicals detected under the present experimental conditions. A reaction pathway directly leading to CH$_3$OH also exists in the CH$_4$ + OH system, namely H-OH substitution reaction, as shown below:

(b)  CH$_4$ + OH → CH$_3$OH + H.        +58.8 kJ mol$^{-1}$ (0.609 eV)

However, this reaction is highly endothermic and has a large activation energy barrier of 44.25 kJ mol$^{-1}$ (1.92 eV), making it unlikely to occur on ASW at 10 K. It is therefore evident that an additional process, beyond reaction (a), must be responsible for methanol formation.

The association reaction between CH$_3$ and OH,

(c)  CH$_3$ + OH → CH$_3$OH,        -367 kJ mol$^{-1}$ (–3.80 eV)

is both exothermic and barrierless, and so is a potential pathway for CH$_3$OH formation under the current experimental conditions (Xu et al. 2007). It should also be noted that there are other exothermic reactions involving CH$_3$ + OH that could occur, as shown below (Xu et al. 2007).

(d)  CH$_3$ + OH → (CH$_3$OH) → H$_2$CO + H$_2$        -293 kJ mol$^{-1}$ (–3.04 eV)
(e)  CH$_3$ + OH → (CH$_3$OH) → *cis*-HCOH + H$_2$    -55.2 kJ mol$^{-1}$ (–0.572 eV)
                              → *trans*-HCOH + H$_2$    -74.1 kJ mol$^{-1}$ (–0.768 eV)

According to the potential energy diagrams presented in Figure 2, reactions (d) and (e) each involve the formation of a CH$_3$OH intermediate. In the case of reaction (d), the barrier from CH$_3$OH to H$_2$CO + H$_2$ is 381 kJ mol$^{-1}$ (3.95 eV), which exceeds the heat released from reaction (c). Thus, reaction (d) would not be expected to proceed at low temperatures even though it is overall exothermic. In the case of reaction (e), the barriers are 368 and 357 kJ mol$^{-1}$ (3.81 and 3.70 eV) for the formation of *cis*- and *trans*-hydroxymethylene (HCOH), respectively. Because these values are almost equal to or lower than the heat of reaction produced by reaction (c), reaction (e) could, in principle, occur even at low temperatures. However, the mass spectra did not show a signal at 163 u corresponding to *cis*-HCOH or *trans*-HCOH, indicating that the internal energy of the intermediate (CH$_3$OH) was readily dissipated to the ice surface. We have recently observed rapid energy dissipation such as this for the reaction OH + CO → HOCO* →

HOCO occurring on ASW (Ishibashi et al. 2024b), in agreement with theoretical predictions (Molpeceres et al. 2023).

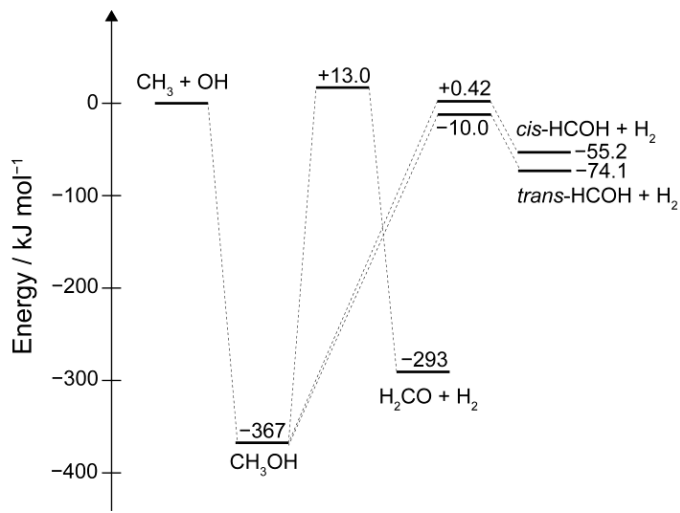

**Figure 2.** Potential energy diagrams (in kJ mol$^{-1}$) of $CH_3$ + OH for reactions (d) and (e). Energies were obtained from the literature (Xu et al. 2007).

It should be noted that, at least on the timescale of the present experiments, the continued generation of $CH_3$ and $CH_3OH$ was not seen after the $CH_4$ deposition was stopped. Hence, reactions forming these two products took place only while $CH_4$ was being supplied. That is, once $CH_4$ and $CH_3$ were thermalized at 10 K, these chemicals were no longer able to encounter the reaction partners, OH radicals. Therefore, at 10 K, the sequential reactions (a) and (c) for generating $CH_3OH$ were likely driven by the non-thermal diffusion of $CH_4$ and $CH_3$ occurring before these molecules reached thermal equilibrium with the ASW surface. Such non-thermal diffusion is also referred to as transient diffusion (Figure 3) or, in the case that the adsorbate is an atom, a hot-atom process. Transient diffusion following adsorbate deposition, especially on well-defined surfaces, has been extensively studied (for a review, see Barth 2000). For example, in an experimental study on the adsorption of Xe on Pt{111} at 4 K, Xe atoms were found to diffuse for hundreds of angstroms before being thermalized (Weiss & Eigler 1992). Transient diffusion following chemical reactions has also been proposed for the energetic processes, such as photochemical reactions, based on molecular dynamics simulations (e.g., Anderson et al. 2006). In the context of nonenergetic reactions, the experiments by Chuang et al. (2016) suggested the possible role of transient diffusion in the formation of COMs during H-atom reactions with mixed CO, $H_2CO$ and $CH_3OH$ ices, although they noted that its contribution could not be directly confirmed from their experiments. However, to the best of our knowledge, in-situ measurements of the transient diffusion of reaction products on a surface have not been reported. Nevertheless, considering the heat released from reaction (a) (55.2 kJ mol$^{-1}$, 0.572 eV), the $CH_3$ produced by this reaction would also be expected to undergo transient diffusion across the ice surface. Indeed, our previous experiments strongly support the occurrence of transient diffusion caused by the heat of reaction (Ishibashi et al. 2024a, 2024b).

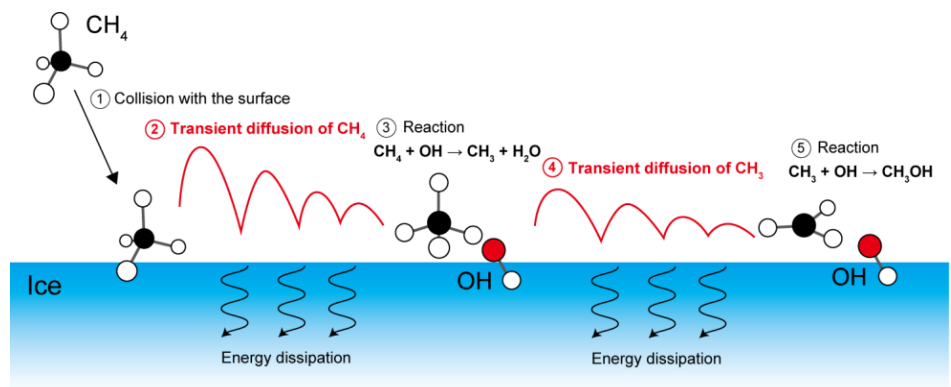

Figure 3. Diagram showing the formation pathway of CH$_3$OH during CH$_4$ deposition onto OH-adsorbed ice via two-step transient diffusion.

Based on the above discussion, the most plausible pathway for the reaction of CH$_4$ and OH radicals at 10 K consists of a sequence of reactions (a) and (c), written as

$$\text{CH}_4{}^* + \text{OH (primary)} \rightarrow \text{CH}_3{}^* + \text{H}_2\text{O} \rightarrow \text{CH}_3 + \text{H}_2\text{O}$$

and

$$\text{CH}_3{}^* + \text{OH (secondary)} \rightarrow \text{CH}_3\text{OH},$$

where CH$_4$* indicates a CH$_4$ molecule that undergoes transient diffusion until reaching thermal equilibrium with the surface at 10 K and CH$_3$* indicates a CH$_3$ radical that also undergoes transient diffusion using the excess energy generated by reaction (a). It should be noted that each CH$_3$OH formation requires two OH radicals based on this reaction process. In addition, only a part of the CH$_4$* molecules are consumed in reaction (a), while the rest become thermalized CH$_4$. As noted earlier, once thermalized at 10 K, CH$_4$ remaining on the ice surface cannot contribute to any further reactions, particularly under low CH$_4$ coverage conditions. Thus, the formation rate of CH$_3{}^*$ can be approximately expressed as the product of the CH$_4{}^*$ flux attacking OH radicals ($F_{\text{CH}_4^*}$), the cross-section for CH$_3{}^*$ formation ($\sigma$), and the surface number density of OH at time $t$ ([OH]$_t$). In the subsequent reaction, CH$_3{}^*$ reacts with a secondary OH radical to produce CH$_3$OH. As with the thermalized CH$_4$, thermalized CH$_3$ cannot diffuse at 10 K and thus remains as detectable CH$_3$ on the surface, with the surface number density [CH$_3$]$_t$. Accordingly, the rate of CH$_3$OH formation can also be expressed as the product of the CH$_3{}^*$ flux ($F_{\text{CH}_3^*}$), the cross-section for CH$_3$OH formation ($\sigma'$), and the surface number density of OH, similar to the CH$_3{}^*$ formation rate. Consequently, assuming that products are not desorbed during the dissipation of excess reaction energy, the rate equations for detectable $CH_3$ and $CH_3OH$ can be written as

$$\frac{d[CH_3]_t}{dt} = \sigma F_{CH_4^*}[OH]_t - \sigma' F_{CH_3^*}[OH]_t, \tag{1}$$

and

$$\frac{d[CH_3OH]_t}{dt} = \sigma' F_{CH_3^*}[OH]_t. \tag{2}$$

Here, $\sigma$ and $\sigma'$ are the cross-sections for $CH_3^*$ and $CH_3OH$ formation for the reactions of transiently diffusing $CH_4^*$ and $CH_3^*$, respectively, with OH, and $F_{CH_4^*}$ and $F_{CH_3^*}$ are the fluxes (i.e., the supply rates) of $CH_4^*$ and $CH_3^*$, respectively. Because the gaseous $CH_4$ was deposited at a constant rate, $F_{CH_4^*}$ was constant over time and thus equal to the $CH_4$ deposition flux, $F_{CH_4}$. Note that this assumes that all deposited $CH_4$ molecules go through the transient $CH_4^*$ state before reacting or undergoing thermalization. In contrast, the value of $F_{CH_3^*}$ changed over time because $F_{CH_3^*}$ is given by the first term of equation (1), $\sigma F_{CH_4^*}[OH]_t$, which contains the time-dependent variable $[OH]_t$. By substituting the first term of equation (1) for $F_{CH_3^*}$ in equations (1) and (2), we obtain these equations as

$$\frac{d[CH_3]_t}{dt} = \sigma F_{CH_4}[OH]_t - \sigma'\sigma F_{CH_4}[OH]_t^2, \tag{3}$$

and

$$\frac{d[CH_3OH]_t}{dt} = \sigma'\sigma F_{CH_4}[OH]_t^2. \tag{4}$$

The derivation of these rete equations has been described in detail in an earlier paper (Ishibashi et al. 2024b).

In the present experiments, the abundance of molecules on the ASW surface was reflected by the pickup intensity, $I_{[X]_t}$, defined as the product of the number density of a given molecule X, $[X]_t$, and the relative pickup efficiency, $\varepsilon'_X$, meaning $I_{[X]_t} = \varepsilon'_X([X]_t/[H_2O]_t)$. Based on this relationship, equations (3) and (4) can be written as

$$\frac{dI_{[CH_3]_t}}{dt} = \frac{\varepsilon'_{CH_3}}{\varepsilon'_{OH}} \sigma F_{CH_4} I_{[OH]_t} - \frac{\varepsilon'_{CH_3}}{\varepsilon'_{OH}{}^2}[H_2O]_t \sigma'\sigma F_{CH_4} I_{[OH]_t}^2, \tag{5}$$

$$\frac{dI_{[CH_3OH]_t}}{dt} = \frac{\varepsilon'_{CH_3OH}}{\varepsilon'_{OH}{}^2}[H_2O]_t \sigma'\sigma F_{CH_4} I_{[OH]_t}^2. \tag{6}$$

Assuming that $[H_2O]_t$ is constant, it was expected that plots of $dI_{[CH_3]_t}/dt$ and $dI_{[CH_3OH]_t}/dt$ as functions of $I_{[OH]_t}$ would obey equations (5) and (6), providing evidence that CH₃OH was indeed produced by the reaction of OH with $CH_3^*$ undergoing transient diffusion using the heat produced by reaction (a). To verify this scenario, we determined the increase rates of the CH₃ and CH₃OH signals near the beginning of CH₄ deposition ($t \sim 0$), meaning $dI_{[CH_3]_{t\sim 0}}/dt$ and $dI_{[CH_3OH]_{t\sim 0}}/dt$, respectively. This was done while varying the initial amount of OH radicals on the ASW prior to CH₄ exposure, $I_{[OH]_{t=0}}$, by changing the duration of UV irradiation.

Figure 4 summarizes the variations in the pickup intensities of CH₃ radical and CH₃OH at 10 K with an initial OH coverage of approximately 0.009. Both intensities increased linearly during the first 200 s of CH₄ deposition. Thus, the early rates of increase, $dI_{[CH_3]_t}/dt|_{t\sim 0}$ and $dI_{[CH_3OH]_t}/dt|_{t\sim 0}$, could be approximated by the slopes of least-squares linear regressions over the time range of 0–200 s, expressed as $dI_{[CH_3]_t}/dt|_{t=0-200}$ and $dI_{[CH_3OH]_t}/dt|_{t=0-200}$, respectively. Note that, over this time span, $I_{[OH]_t}$ was nearly constant, meaning that pseudo first order reaction conditions were achieved.

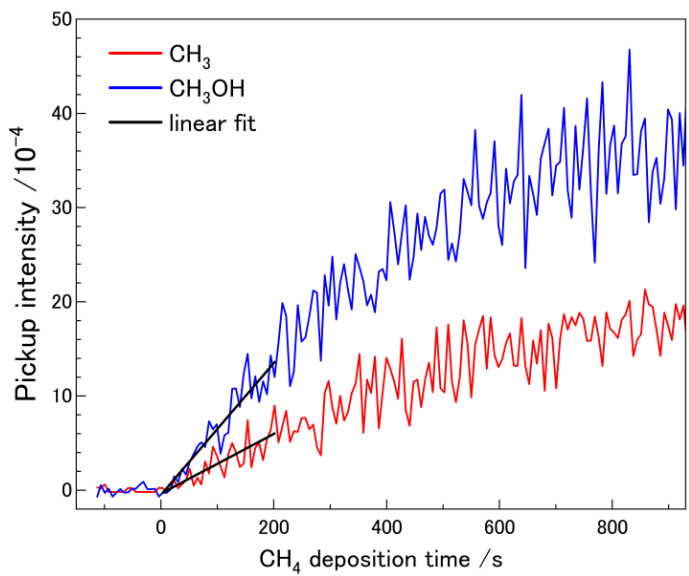

**Figure 4.** CH₃ radical (red) and CH₃OH (blue) pickup intensities during CH₄ deposition on UV-irradiated ASW surface, as functions of time. The initial coverage of OH radicals was approximately 0.009, corresponding to 9 minutes of UV irradiation. The solid black lines represent linear fitting results using the data for 0–200 s.

The signal increase rates were determined for various initial OH coverages and these data are plotted in Figure 5. These plots could be fitted very well using equations (5) and (6). These results support our proposed CH₃OH formation mechanism. Defining the coefficients for the first and second terms of equation (5) as $A$ and $B$, respectively, and the coefficient for the term of equation (6) as $C$, these regressions yielded $A = (3.5 \pm 0.5) \times 10^{-4}$ s$^{-1}$, $B = -(7.7 \pm 4.2) \times 10^{-7}$, and $C = (4.9 \pm 0.2) \times 10^{-6}$. Based on the present definitions, the ratio $-B/C$ indicates that the ratio of the pickup efficiencies, $\varepsilon_{CH_3}/\varepsilon_{CH_3OH}$, was 0.16, assuming the desorption of CH₃ and CH₃OH upon their formation (chemical desorption) was negligible. Therefore, the actual number density ratio ( $[CH_3]_t$ / $[CH_3OH]_t$ ) could be represented by $I_{[CH_3]_t}/0.16 I_{[CH_3OH]_t}$. Using this relationship, the fraction of CH₃* consumed for CH₃OH formation was estimated as

$$\frac{[CH_3OH]_t}{[CH_3OH]_t+[CH_3]_t} = \frac{1}{1+I_{[CH_3]_t}/0.16 I_{[CH_3OH]_t}}, \quad (7)$$

where the sum ($[CH_3OH]_t + [CH_3]_t$) corresponds to the amount of CH₃* produced. Using this equation, the fraction of CH₃* consumption could be calculated from pickup intensities $I_{[CH_3]_t}$ and $I_{[CH_3OH]_t}$. As an example, the data at $t = 200$ s in Figure 4, $I_{[CH_3]_{t=200}} = 6$ and $I_{[CH_3OH]_{t=200}} = 13.6$, gives a fraction of 27%. Similar analyses were carried out for various initial OH coverages and the percentage of consumed CH₃* is plotted as a function of the OH coverage in Figure 6.

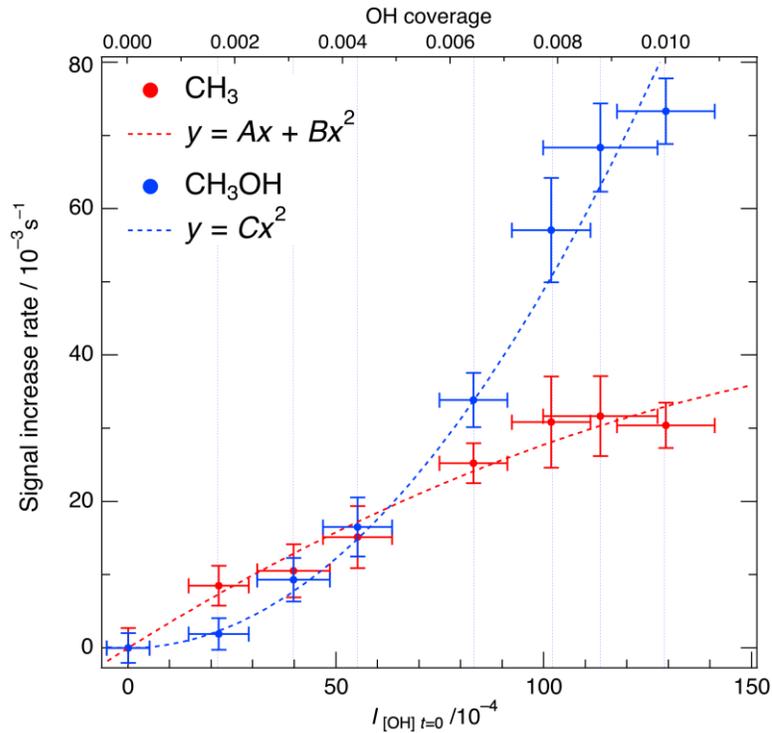

**Figure 5.** Signal increase rates for $CH_3$ (red dots) and $CH_3OH$ (blue dots) as functions of the OH pickup intensity at the beginning of $CH_4$ deposition, $I_{[OH]_{t=0}}$. The initial OH coverage is given on the upper axis. These values were calculated assuming that the OH coverage after 10 minutes of UV irradiation was approximately 0.01 (see Section 2). Error bars indicate statistical errors. The red and blue dashed lines represent the fitting curves based on equations (5) and (6), respectively, $y = Ax + Bx^2$ and $y = Cx^2$.

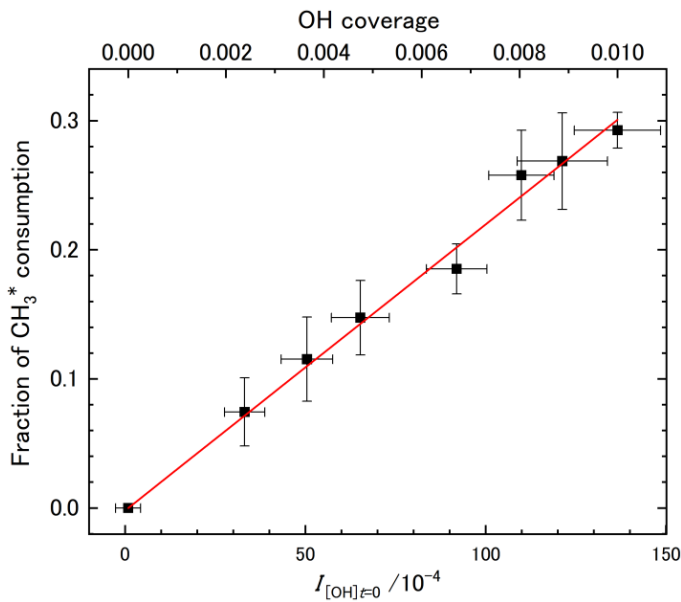

**Figure 6.** Fraction of $CH_3^*$ consumed during $CH_3OH$ formation (equation (7)) as function of OH pickup intensity at beginning of $CH_4$ deposition, $I_{[OH]_{t=0}}$. The initial OH coverage is given on the upper axis. These values were calculated assuming that the OH coverage after 10 minutes of UV irradiation was approximately 0.01 (see Section 2). The red line indicates the linear fit to the data.

As shown in Figure 6, it is clear that the consumed $CH_3^*$ fraction was proportional to the initial OH coverage. This result is not unexpected, assuming that the transient diffusion length was limited. That is, the probability that a $CH_3^*$ molecule undergoing diffusion encountered an OH radical before being thermalized should be proportional to the OH number density. It is of note that quite a large percentage (approximately 25%) was consumed even when the OH coverage was as low as 0.01. On actual ice dust, the species reacting with $CH_3^*$ need not be an OH radical but rather could be a number of other reactive radicals, such as HCO or $NH_2$. Hence, when the total number density of all such radicals could be high enough, a large percentage of the $CH_3^*$ produced by the $CH_4$ + OH reaction would contribute to COMs formation via radical–radical reactions even at 10 K.

*3.2 Effect of temperature on CH₃OH formation on ASW*

The effects of temperature on reactions (a) and (c) were evaluated by monitoring the pickup intensities for $CH_3$, OH, and $CH_3OH$ during $CH_4$ deposition at temperatures ranging from 10 to 60 K. In these experiments, 60 K was selected as an upper limit because OH diffusion on ASW becomes pronounced above this temperature, leading to the loss of OH radicals via OH–OH recombination to produce $H_2O_2$ (Miyazaki et al. 2022). Figure 7 shows plots of the pickup intensities for OH, $CH_4$, $CH_3$, and $CH_3OH$ as functions of the $CH_4$ deposition time, $t$.

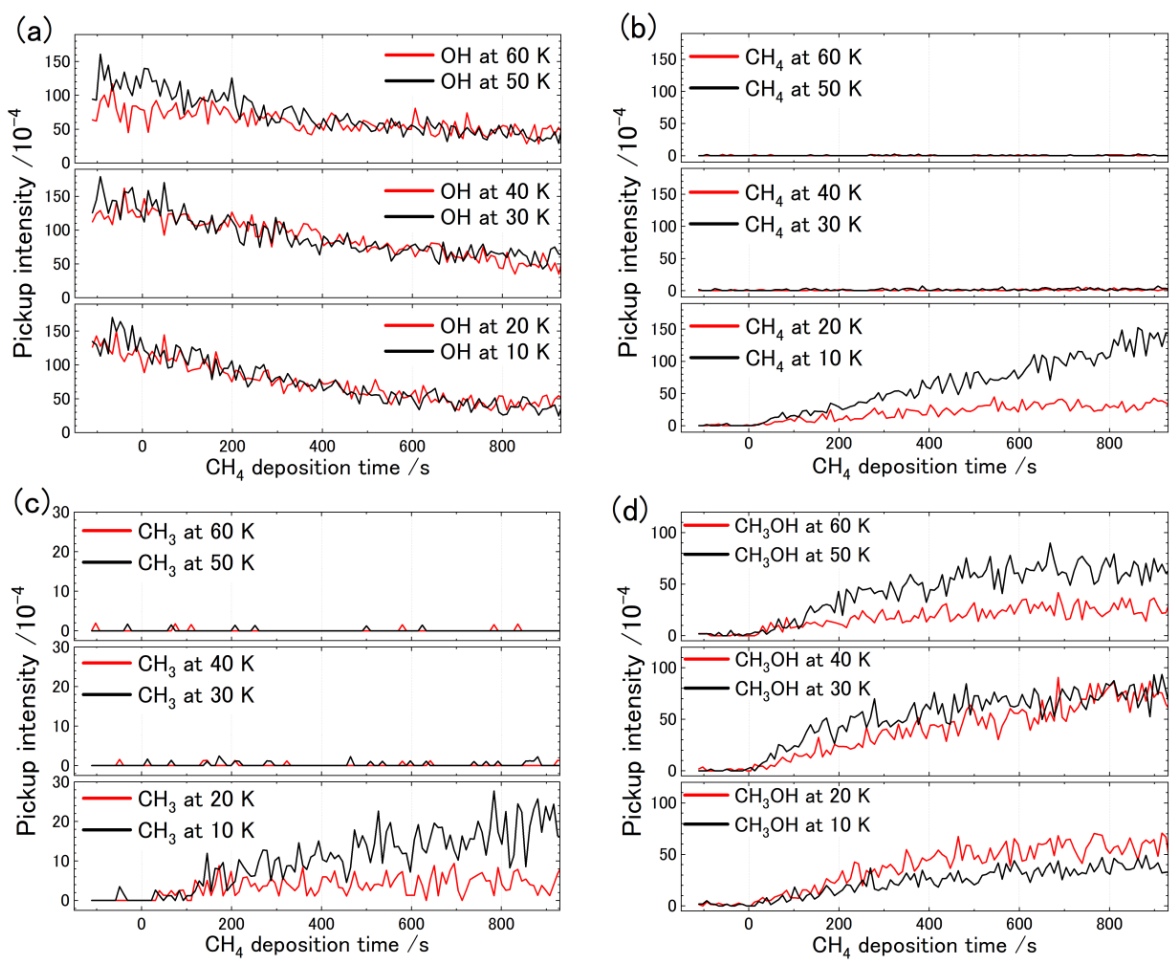

**Figure 7.** Pickup intensities for (a) OH, (b) $CH_4$, (c) $CH_3$, and (d) $CH_3OH$ as functions of time at different ASW temperatures. All data were obtained using UV-irradiated ASW with an OH coverage of 0.01. UV irradiation was conducted at 30 K for measurements at 10, 20, and 30 K but at 40, 50, and 60 K for measurements at those same temperatures.

As can be seen from Figure 7(a), the OH radical intensity decreased with increases in the $CH_4$ deposition time at all temperatures, indicating that the reaction of OH with $CH_4$ occurred regardless of the surface temperatures. Moreover, the signal decay rates were similar across all temperatures. The slightly lower initial OH intensity at 60 K can likely be attributed to some OH diffusion events during UV irradiation, leading to OH–OH recombination. As shown in Figures 7(b) and 7(c), the pickup intensities for $CH_4$ and for $CH_3$ radicals became negligible above 30 K. The absence of $CH_4$ above 40 K is reasonable when considering that $CH_4$ desorbs from the surface of ASW in this temperature range (Smith et al. 2016). Consequently, the residence time of $CH_4$ on the ASW surface would be too short for accumulation and detection. We concluded that the absence of a $CH_4$ signal at 30 K may be attributed to the significantly lower $CH_4$ pickup efficiency at that temperature due to the diffusion of $CH_4$ to sites where the $Cs^+$ ions cannot access and pick up $CH_4$. Indeed, Furuya *et al*. (2022) previously demonstrated that the diffusion of $CH_4$ during its deposition on compact ice above 20 K is sufficient to allow the formation of crystalline $CH_4$. In the work reported herein, $CH_3OH$ formation was observed even at temperatures above 30 K, as shown in Figure 7(d). The intensity of the $CH_3OH$ signal reached a maximum at approximately 30 K and remained relatively high up to 50 K compared with the value at 10 K. Because these signal variations could have resulted from the effects of temperature on the pickup efficiency (Ishibashi et al. 2021), we also conducted a control experiment to evaluate the temperature dependence of the $CH_3OH$ pickup efficiency over the range of 10 to 60 K.

Figure 8 shows the temperature dependence of $CH_3OH$ pickup intensity on pure ASW. After $CH_3OH$ injection onto the ASW, we increased the temperature from 10 to 60 K. The blue trace represents the smoothed version of the raw $CH_3OH$ pickup intensity data shown in gray, while the red trace indicates the ASW temperature. From this figure, it is clear that the $CH_3OH$ pickup intensity remained essentially constant across this temperature range. Because $CH_3OH$ did not desorb within this temperature range, the surface number density of this molecule was likely unchanged over time. This constant pickup signal confirms that the $CH_3OH$ pickup efficiency was temperature independent between 10 and 60 K. Therefore, it appears that the variations in the $CH_3OH$ signal seen in Figure 7(d) must reflect changes in the number density of $CH_3OH$ on the ASW surface. These results demonstrate that both $CH_4$ and $CH_3$ radicals were able to rapidly react with OH radicals to generate $CH_3OH$ prior to undergoing thermal desorption.

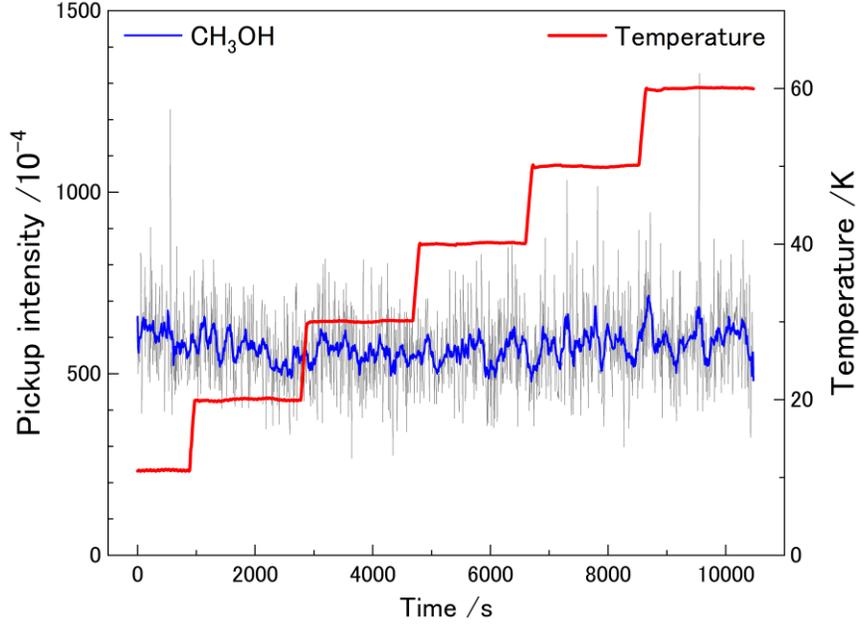

**Figure 8.** Effect of temperature on CH$_3$OH pickup intensity. The gray and blue traces indicate raw and smoothed data, respectively. During this measurement, the temperature of the ASW with deposited CH$_3$OH was increased from 10 to 60 K in a stepwise manner as indicated by the red line. Prior to acquiring data, CH$_3$OH was deposited on the ASW at 60 K after which the sample was cooled to 10 K. The resulting surface coverage by CH$_3$OH was approximately 0.03.

At an OH coverage of 0.01, a CH$_4$ molecule must reach 100 adsorption sites, on average, to encounter an OH radical. Moreover, because the reaction of CH$_4$ + OH → CH$_3$ + H$_2$O is thought to occur via quantum mechanical tunneling at low temperatures, a number of collisions with OH radicals would be required for a reaction to occur. The time required to visit 100 adsorption sites (the diffusion time, $t_{\text{diff}}$) can be calculated assuming thermal diffusion. The value of $t_{\text{diff}}$ can be estimated according to the equations

$$D = D_0 exp\left(-\frac{E_{\text{diff}}}{k_{\text{B}}T}\right), \tag{9}$$

and

$$t_{\text{diff}} = 100S/D, \tag{10}$$

where $D$ and $D_0$ are the diffusion coefficient and a pre-exponential factor, respectively, $E_{\text{diff}}$ is the activation energy for diffusion, $k_{\text{B}}$ is Boltzmann's constant, and $S$ is the area of a unit adsorption site (1 × 10$^{-15}$ cm$^2$). Using a typical $D_0$ value of 9 × 10$^{-4}$ cm$^2$ s$^{-1}$ and an experimentally derived $E_{\text{diff}}$ value of 200 K (Furuya et al. 2022), the $t_{\text{diff}}$ at 30 and 60 K were estimated to be 9 × 10$^{-8}$ and 3 × 10$^{-9}$ s, respectively. Because transient diffusion promoted surface diffusion, it is quite likely that both adsorbed CH$_4$* and produced $CH_3^*$ migrated over significant areas on the ice surface, leading to $CH_3OH$ formation. Therefore, the significant $CH_3OH$ formation observed above 30 K is attributed to reactions (a) and (c), even though very low amounts of $CH_4$ and $CH_3$ were found on the ASW. The enhanced $CH_3OH$ formation at 30–40 K can possibly be attributed to an increase in the probability that $CH_4$ or $CH_3$ encountered OH based on extended diffusion lengths. These longer lengths are achieved by the contribution of both transient and thermal diffusion processes within the residence times of the two species. In fact, the activation of thermal diffusion of $CH_3$ radicals above 15-20 K has been reported in the context of methylamine formation on ice at low temperatures (Iguchi et al. 2025). However, the increased $CH_3OH$ formation did not lead to significant variation in the intensity of the OH radical signal (Figure 7(a)). This result indicates that the yield of $CH_3OH$ relative to the amount of $CH_3$ produced, i.e., the fraction of $CH_3^*$ consumption, increased at high temperatures, probably because of an extended diffusion length of $CH_3^*$. The reduced $CH_3OH$ signal at 60 K could have been at least partly a consequence of the lower OH abundance caused by $H_2O_2$ formation. Moreover, the more rapid thermal desorption of $CH_4$ and $CH_3$ at high temperatures would lead to shorter diffusion lengths.

## 4. ASTROPHYSICAL IMPLICATIONS

It is known that OH radicals can be generated on ice dust surfaces by several processes, including the association of atomic oxygen and hydrogen atoms (Dulieu et al. 2010). These processes occur by the successive hydrogenation of $O_2$ (Miyauchi et al. 2008; Ioppolo et al. 2008) and the photodissociation of $H_2O$. Therefore, OH radicals are expected to be abundant on such surfaces. The abundance of solid $CH_4$ relative to solid $H_2O$ has been reported as <3% toward the background stars CK2 and Elias 16 (Knez et al. 2005), and as 2% and 2.5% toward J110621 and NIR 38, respectively (McClure et al. 2023). Furthermore, the successive hydrogenation of carbon atoms (Qasim et al. 2020; Tsuge et al. 2024) and the reaction of $CH_3 + H_2 \rightarrow CH_4 + H$ (Lamberts et al. 2022) have been experimentally confirmed as $CH_4$ formation processes on ice dust. Therefore, the reaction between $CH_4$ and OH is likely to occur on ice dust in molecular clouds.

Qasim *et al*. (2018) reported $CH_3OH$ generation following the co-deposition of $CH_4$, $O_2$, and H atoms onto a substrate at temperatures of 10-20 K over durations of 6–12 h. Their work confirmed $CH_3OH$ formation in a $CH_4$-$O_2$-$H_2$ matrix, the thickness of which was estimated to be several hundred monolayers based on the duration of deposition and the fluxes of co-deposited beams. The pathway for $CH_3OH$ production under such conditions would be similar to that confirmed in the present study, although $CH_3OH$ was identified as the sole final product.

The present study confirms that CH₃OH formation on ASW occurred via two elementary reactions. The first of these, $CH_4 + OH \rightarrow CH_3 + H_2O$, took place even at 10 K and was facilitated by the transient diffusion of $CH_4$ and hydrogen abstraction in association with quantum tunneling. The subsequent reaction of a $CH_3$ radical with a second OH radical, $CH_3 + OH \rightarrow CH_3OH$, involved the transient diffusion of a $CH_3$ radical utilizing the heat of the first reaction. This work demonstrates that these sequential reactions could potentially form $CH_3OH$ on ice dust. Furthermore, a significant proportion (approximately 25%) of the $CH_3$ radicals produced by the first reaction are thought to have subsequently contributed to $CH_3OH$ formation. Hence, transient diffusion promoted by the heat of the initial reaction clearly played an important role in driving subsequent chemical reactions, even with the low reactant coverages expected to occur in astrophysical environments. It appears that diffusive reactions between radicals heavier than hydrogen atoms can occur even at 10 K through this mechanism. Consequently, other $CH_3$ formation pathways with comparable or greater heats of reaction, such as $CH + H_2 \rightarrow CH_3$ (-443 kJ mol⁻¹, –4.59 eV) or $CH_2 + H \rightarrow CH_3$ (-457 kJ mol⁻¹, –4.74 eV) (Blitz et al. 2023), could also produce $CH_3OH$ on actual ice dust surfaces. The generation of $CH_3$ via the reaction of $CH_2 + H_2 \rightarrow CH_3 + H$ on low-temperature solid surfaces has been previously suggested (Lamberts et al. 2022). However, the contribution of this process to $CH_3OH$ formation might be limited due to the high activation barrier (42.3 kJ mol⁻¹, 0.438 eV) (Lu et al. 2010) and relatively low released heat (-23.4 kJ mol⁻¹, -0.243 eV).

The effect of temperature on $CH_3OH$ formation, as determined in this study, indicates that the $CH_3OH$ formation can proceed efficiently even at 60 K, at which temperature the parent molecules (i.e., $CH_4$ and $CH_3$ radicals) will not be adsorbed stably on the ASW surface. This finding suggests that radical reactions without activation energies should, in general, be considered possible even if the dust temperature exceeds the desorption temperature of the reactants included in typical astrochemical models.

This study also shows that the transient diffusion of $CH_3$ radicals on ASW surfaces, driven by the heat released from the reactions above, may contribute significantly to the formation of COMs such as $CH_3CHO$ and $CH_3CH_2OH$ that have been recently identified on interstellar ice (Rocha et al. 2024). The formation of these species is likely to occur by reactions involving $CH_3$ radicals and reaction partners such as HCO or $CH_2OH$, both of which would be relatively abundant on interstellar ice surfaces.

## 5. CONCLUSIONS

The present experimental results clearly demonstrate that $CH_3OH$ can be produced by radical–radical reactions on ice dust surfaces. This mechanism is an alternative pathway to the hydrogenation of CO. These findings indicate that reactions between heavier radicals can contribute to the formation of COMs via transient diffusion, driven by the heat released during exothermic reactions, on an ice dust surface at 10 K. Notably, the present process occurs even within a temperature range over which hydrogen atoms readily desorb and the thermal diffusion of heavier species is inefficient. Molecular formation within the ice matrix is also plausible route for methanol production. The radical-radical pathway $CH_3 + OH \rightarrow CH_3OH$ has been reported inside ice under high energy electron irradiation of $CH_4$-$H_2O$ mixed ice (Bergantini et al. 2017). Another viable mechanism involves oxygen insertion into methane, $O\,(^1D) + CH_4 \rightarrow CH_3OH$, as demonstrated in UV irradiated $O_2$-$CH_4$ mixed ice (Jennifer et al. 2018). Although assessing the relative efficiencies of these processes is challenging, radical–radical reactions mediated by transient diffusion are expected to play a role and this should be incorporated into astrochemical models. Such models could be further refined by quantitative constraints on transient diffusion lengths for specific molecular species along with internal energies. For this purpose, it will be crucial to perform experimental work under conditions with low reactant coverages that realistically approximate ice dust surfaces. The use of high-sensitivity analytical techniques such as the $Cs^+$ pickup method, which is able to detect small amounts of molecules on ASW as demonstrated in the present work, would be helpful in such investigations.

## ACKNOWLEDGEMENTS

This work was supported in part by JSPS KAKENHI grant numbers JP25K17453 (A.I.), JP25K07377 (H.H.), JP 24K00686 (M.T.), JP22H00159 (N.W.), and JP20H05847.